\documentstyle[aps,epsfig]{revtex}
\begin{document}

\title{ On the extension of the Bethe-Weizs\"acker mass formula to light nuclei }
\author{D.N. Basu\thanks{E-mail:dnb@veccal.ernet.in}}
\address{Variable  Energy  Cyclotron  Centre,  1/AF Bidhan Nagar,
Kolkata 700 064, India}
\date{\today }
\maketitle
\begin{abstract}

      Some general features of the Bethe-Weizs\"acker mass formula recently extended to light nuclei have been explored. Though this formula  improves fits to the properties of light nuclei and it does seem to work well in delineating the positions of all old and new magic numbers found in that region, yet it is not well tuned for predicting finer details. The mass predictions have also been found to be less accurate compared to those by the macroscopic-microscopic calculations. It is concluded that such semi-empirical mass formulae can not be a substitute for more fundamental mass formulae having its origin based upon the basic nucleon-nucleon effective interaction. 

\noindent 

\end{abstract}



      In a recent article \cite{r1} the Bethe-Weizs\"acker (BW) mass formula has been extended to light nuclei and some new shell closures have been identified. This modified BW mass formula explains the shapes of the binding energy versus neutron number curves of most of the elements from $Li$ to $Bi$. For heavier nuclei this modified BW mass formula approaches the old BW mass formula. This modified BW formula suggests additional stabilities for few light nuclei at neutron numbers N=16 (Z=7,8), N=14 (Z=7-10), Z=14 (N=13-19), N=6 (Z=3-8) and loss of magicity for nuclei with neutron numbers N=8 (Z=4) and N=20 (Z=12-17), most of which were suggested earlier from the systematics. The magicity at neutron number N=6 (Z=3-8) suggested by the new formula is also supported by the experimental r.m.s. radii values that show a quenching at N=6 (Z=3-8). The new formula supports already known $^{32}Ne$, $^{35}Na$, $^{38}Mg$ and $^{41}Al$ as the last bound isotopes of neon, sodium, magnesium and aluminium, respectively. The existing mass formulae \cite{r2,r3} support the existence of old magic numbers but fail to explain the general trend of the binding energy versus neutron number curves for several light nuclei near the drip lines and evidence of new magic number at neutron number N=16. Wherever there is a signature of shell closure in experimental data, the BW as well as the new modified BW formula both show marked deviations delineating clearly the positions of old magic numbers at 2, 8, 20, 28, 50, 82 and 126. 

      The binding energy B(A,Z) of any nucleus of mass number A and atomic number Z was obtained from a phenomenological search and was given by a more genaralized BW formula \cite{r1} 

\begin{equation}
 B(A,Z) = a_vA-a_sA^{2/3}-a_cZ(Z-1)/A^{1/3}-a_{sym}(A-2Z)^2/[(1+e^{-A/k})A]+\delta_{new},
\label{seqn1}
\end{equation}
\noindent
where the asymmetry term and the pairing term had been modified to obtain great improvements to the fits of the binding enegy versus neutron number curves of the light nuclei. Fitted value of the asymmetry term modifying constant $k=17$, while the modified pairing term $\delta_{new}$ can be expresssed as  

\begin{equation}
 \delta_{new}=(1-e^{-A/c})\delta~~~~where~~~~c=30,
\label{seqn2}
\end{equation}
\noindent
and $\delta$ being the old pairing term is given by

\begin{eqnarray}
  \delta=&&a_pA^{-1/2}~for~even~N-even~Z,\nonumber\\
         =&&-a_pA^{-1/2}~for~ odd~N-odd~Z,\nonumber\\
         =&&0~for~odd~A,\nonumber\\
\label{seqn3}
\end{eqnarray} 
\noindent
while the other constants

\begin{equation}
 a_v=15.79~MeV,~a_s=18.34~MeV,~a_c=0.71~MeV,~a_{sym}=23.21~MeV~and~a_p=12~MeV,
\label{seqn4}
\end{equation}
\noindent
remaining the same as the corresponding values of the constants for the old BW formula. Such modifications, however, do not alter significantly the results for heavier nuclei. 

      In order to investigate the features of this newly extended Bethe-Weizs\"acker mass formula, the one and the two proton separation energies, the one and the two neutron separation energies and the atomic mass excesses have been calculated using the formula given by Eq.(1) with the $\delta_{new}$ quantity described by Eq.(2). The best fit values used for the new constants are $k=17$ and $c=30$ while the values of the older constants used for calculating the binding energies are given in Eq.(4). The one and two proton separation energies $S_{1p}$ and $S_{2p}$ defined as the energies required to remove one proton and two protons from a nucleus are given by

\begin{equation}
 S_{1p} = B(A,Z) - B(A-1,Z-1)~~~~and~~~~S_{2p} = B(A,Z) - B(A-2,Z-2)
\label{seqn5}
\end{equation}   
\noindent
respectively, while the one neutron and two neutron separation energies $S_{1n}$ and $S_{2n}$ defined as the energies required to remove one neutron and two neutrons from a nucleus are given by

\begin{equation}
 S_{1n} = B(A,Z) - B(A-1,Z)~~~~and~~~~S_{2n} = B(A,Z) - B(A-2,Z)
\label{seqn6}
\end{equation}   
\noindent
respectively. The reason that the atomic mass is considered rather than the nuclear mass is that historically, the former has been the actual experimentally measured quantity, whereas  the latter is less accurate bacause its extraction requires a knowledge of binding energy of the Z atomic electrons. However, recent developments now allow the nuclear masses to be measured directly \cite{r4}. For those applications where it is necessary to know the  actual mass of the nucleus itself, its value (in $MeV$) can be found from the atomic mass excess or from nuclear binding energy by use of the relationship

\begin{eqnarray}
 M_{nucleus}(A,Z) =&& A u + \Delta M_{A,Z} - Zm_e + a_{el} Z^{2.39} \nonumber\\
                           =&& Z m_p + (A-Z) m_n - B(A,Z),~by~definition~of~nuclear~binding~energy~B(A,Z)
\label{seqn7}
\end{eqnarray}   
\noindent
where $m_p$, $m_n$ and $m_e$ are the masses of proton, neutron and electron respectively, the atomic mass unit u is 1/12 the mass of $^{12}C$ atom, $\Delta M_{A,Z}$ is the atomic mass excess of an atom of mass number A and atomic number Z and the electronic binding energy constant $a_{el} = 1.433 \times 10^{-5}$ MeV. Hence from Eqs.(7) the atomic mass excess is given by 

\begin{eqnarray}
 \Delta M_{A,Z} =&& Z m_p + (A-Z) m_n + Zm_e - A u - a_{el} Z^{2.39} - B(A,Z)  \nonumber\\
                       =&& Z (m_p + m_e - u) + (A-Z) (m_n - u) - a_{el} Z^{2.39} - B(A,Z) \nonumber\\
                       =&& Z \Delta m_H  + (A-Z) \Delta m_n - a_{el} Z^{2.39} - B(A,Z)
\label{seqn8}
\end{eqnarray}
\noindent
where $\Delta m_H = m_p + m_e - u$ = 7.289034 MeV and $\Delta m_n = m_n - u$ = 8.071431 MeV.

      Though such modifications certainly cause impovements over earlier semi-empirical mass formulae, but it is still beset with many difficulties. The new formula suggests that the $^{31}F$ nucleus is unstable while experimental evidence of its existence has been found. However, similar instability was predicted by other models also. The formula wrongly predicts $^5Li$, $^6Be$, $^{15}F$ and $^{16}Ne$ as stable \cite{r5}, and puts the neutron drip line at $^{17}B$, $^{20}C$ and $^{26}O$ \cite{r5} whereas $^{19}B$, $^{22}C$ are in fact also stable, but $^{25,26}O$ are not. While it may be used to provide gross features like predicting approximate positions of the drip lines, it must be realised that a semi-empirical formula such as the Bethe-Weizs\"acker mass formula can not be well tuned for predicting finer details. 

\begin{table}
\caption{One and two proton separation energies for proton rich and one and  two neutron separation energies for neutron rich nuclei predicted by the modified Bethe-Weizs\"acker mass formula}
\begin{tabular}{cccccc}
Nuclei&$S_{1p}$&$S_{2p}$&Nuclei&$S_{1n}$&$S_{2n}$  \\ 
  &$MeV$&$MeV$& &$MeV$&$MeV$ \\ \hline

  $^5Li$&3.37&15.16&$^{17}B$&1.00&0.56   \\ 
  $^6Be$&1.12&4.49&$^{20}C$&1.03&0.42  \\ 
  $^{15}F$&0.11&4.75&$^{26}O$&0.95&-0.01  \\ 
  $^{16}Ne$&0.65&0.76&$^{31}F$&-0.38&-2.91 \\
 
\end{tabular} 

\end{table}
\nopagebreak

      The calculated theoretical atomic mass excesses using the macroscopic-microscopic model \cite{r3} and the semi-empirical modified Bethe-Weizs\"acker (BW) mass formula \cite{r1} denoted by $\Delta M_{MS}$ and $\Delta M_{BW}$ respectively, have been calculated using Eq.(8) while the atomic mass excesses denoted by $\Delta M_{Ex}$ are the experimentally measured quantities. The chi-square per degree of freedom $\chi^2/F$ can be used for a relative comparison between different theoretical models which is defined as 

\begin{equation}
 \chi^2/F =  (1/N) \sum [ ( \Delta M_{Th} - \Delta M_{Ex} ) / \Delta M_{Ex} ]^2
\label{seqn9}
\end{equation}   
\noindent
where $\Delta M_{Th}$ is the theoretically calculated atomic mass excess and the summation extends to $N$ data points for which experimental atomic mass excesses are known. The $\chi^2/F$ calculated for the semi-empirical Bethe-Weizs\"acker (BW) mass formula using 1742 experimental atomic mass excesses has been found to be 12.936. The $\chi^2/F$ for the macroscopic-microscopic model using 1727 experimental atomic mass excesses has been found to be 1.821. Only one point corresponding to the $^{12}C$ has been excluded from the summation in both the calculations for $\chi^2/F$ because the experimental atomic mass excess for the $^{12}C$ atom is, by definition, equal zero and can not be accomodated inside the summation given by Eq.(9). It is evident that the macroscopic-microscopic model \cite{r3} which uses momentum and density dependent Seyler-Blanchard \cite{r6} type effective interaction provide better mass estimates than the semi-empirical modified Bethe-Weizs\"acker (BW) mass formula.   
 
      To conclude, the extension of Bethe-Weizs\"acker mass formula to light nuclei is beset with some difficulties. There are predictions about the stability of some light nuclei which are not in agreement with the experimental observations. Mass estimates also show clear superiority for more fundamental mass formulae having its origin based upon the basic nucleon-nucleon effective interaction. The need for improvements of mass formula having its origin based upon the basic nucleon-nucleon effective interaction is stressed rather than modifying semi-empirical mass formula, such as the Bethe-Weizs\"acker, which can not be tuned to reproduce finer details. Ideally, a density dependent effective interaction \cite{r7} which provides unified descriptions of the nuclear scattering \cite{r8,r9}, cluster radioactivity \cite{r10} and nuclear matter \cite{r11} may be used for further improvements in the macroscopic-microscopic model.


\end{document}